\documentclass[12pt]{article}
\pdfoutput=1

\usepackage{amsmath}
\usepackage{amsfonts}
\usepackage{amscd}
\usepackage{amsthm}
\usepackage{setspace}

\usepackage{graphicx}
\usepackage{authblk}
\usepackage{caption}
\usepackage{ytableau}
\usepackage{mathtools}

\usepackage[OT2,T1]{fontenc}
\DeclareSymbolFont{cyrletters}{OT2}{wncyr}{m}{n}
\DeclareMathSymbol{\Sha}{\mathalpha}{cyrletters}{"58}

\def\Z{\mathbb{Z}}

\def\C{\mathbb{C}}
\def\P{\mathbb{P}}

\begin{document}

\begin{titlepage}

\begin{flushright}
KEK-TH-2130
\end{flushright}

\vskip 1cm

\begin{center}

{\bf \Large Discrete gauge groups in certain F-theory models \\
 \vspace{0.5cm}
in six dimensions}

\vskip 1.2cm

Yusuke Kimura$^1$ 
\vskip 0.4cm
{\it $^1$KEK Theory Center, Institute of Particle and Nuclear Studies, KEK, \\ 1-1 Oho, Tsukuba, Ibaraki 305-0801, Japan}
\vskip 0.4cm
E-mail: kimurayu@post.kek.jp

\vskip 2cm
\abstract{We construct six-dimensional (6D) F-theory models in which discrete $\Z_5, \Z_4, \Z_3,$ and $\Z_2$ gauge symmetries arise. We demonstrate that a special family of ``Fano 3-folds'' is a useful tool for constructing the aforementioned models. The geometry of Fano 3-folds in the constructions of models can be useful for understanding discrete gauge symmetries in 6D F-theory compactifications. We argue that the constructions of the aforementioned models are applicable to Calabi--Yau genus-one fibrations over any base space, except models with a discrete $\Z_5$ gauge group. We construct 6D F-theory models with a discrete $\Z_5$ gauge group over the del Pezzo surfaces, as well as over $\P^1\times\P^1$ and $\P^2$. We also discuss some applications to four-dimensional F-theory models with discrete gauge symmetries.}  

\end{center}
\end{titlepage}

\tableofcontents
\section{Introduction}

\par The aim of this study is to construct and discuss six-dimensional (6D) F-theory models with discrete gauge groups. F-theory \cite{Vaf, MV1, MV2} is compactified on a space that admits a genus-one fibration, and there are two types of situations: where this fibration structure admits a global section and where it does not. The complex structure of a genus-one fiber of this fibration structure is identified with the axio-dilation. This formulation enables the axio-dilaton to admit an $SL_2(\Z)$ monodromy. F-theory compactifications on elliptic fibrations with a global section have previously been analyzed \cite{MorrisonPark, MPW, BGK, BMPWsection, CKP, BGK1306, CGKP, CKP1307, CKPS, AL, EKY1410, LSW, CKPT, CGKPS, MP2, BMW2017, CL2017, BMW1706, EKY1712, KimuraMizoguchi, Kimura1802, LRW2018, MizTani2018, CMPV1811, TT2019, Kimura1903}. Furthermore, the physical interpretation of F-theory compactification on a genus-one fibration without a section was recently discussed in \cite{BM, MTsection}. It was argued \cite{MTsection} that a discrete gauge symmetry \footnote{Recent progress of discrete gauge groups can be found, for example, in \cite{KNPRR, ACKO, BS, HSsums, CIM, BISU, ISU, BCMRU, BCMU, MRV, HS, BRU, KKLM, HS2, GPR}.} arises in the context of F-theory compactification on a genus-one fibration lacking a global section. F-theory models on genus-one fibration lacking a global section have recently been studied, for example, in \cite{BM, MTsection, AGGK, KMOPR, GGK, MPTW, MPTW2, BGKintfiber, CDKPP, LMTW, K, K2, ORS1604, KCY4, CGP, Kdisc, Kimura1801, AGGO1801, Kimura1806, TasilectWeigand, CLLO, TasilectCL, HT, Kimura1810, Kimura1902} \footnote{\cite{BEFNQ, BDHKMMS} studied F-theory compactifications on genus-one fibrations without a section.}. A genus-one fibration without a global section still admits a multisection, and when the multisection has degree $n$, it can be interpreted as an $n$-fold cover of the base space of the genus-one fibration. The special situation in which this $n$-fold cover splits into $n$ disjoint sheets of global sections, yields an interpretation as an elliptic fibration with $n$ independent global sections \cite{MTsection}. From this viewpoint, the moduli of F-theory models on genus-one fibrations contain the moduli of F-theory models on elliptic fibrations with multiple sections as a subset \cite{MTsection}. 
\par When a fibration structure associated with a genus-one fibration exists, called the ``Jacobian fibration,'' taking the Jacobian fibration \footnote{\cite{Cas} discusses a construction of the Jacobian of an elliptic curve.} of the genus-one fibration yields an elliptic fibration with a global section. This operation of ``taking the Jacobian fibration'' relates the moduli of genus-one fibrations without a section to those of the Weierstrass models of F-theory on elliptic fibrations with a section \cite{MTsection}. The $\tau$ function of a Calabi--Yau genus-one fibration $Y$ and that of the Jacobian fibration $J(Y)$ are identical.
\par In the moduli of F-theory models without a section, the process in which the multisection of a genus-one fibration splits into multiple global sections and the genus-one fibration becomes an elliptic fibration with a global section can be viewed as the reverse of ``Higgsing,'' where a $U(1)$ gauge group breaks and the discrete gauge group remains \cite{MTsection}. This represents the physical interpretation of F-theory on genus-one fibrations without a global section discussed in \cite{MTsection}, where a discrete $\Z_n$ gauge symmetry arises in F-theory on a genus-one fibration \footnote{Discrete gauge group forming in F-theory compactification is given as the discrete part of the ``Tate--Shafarevich group'', $\Sha$, of the Jacobian fibration of the compactification (Calabi--Yau) space \cite{MTsection}. When one considers F-theory compactification on genus-one fibered Calabi--Yau manifold $Y$, one can consider the ``Tate--Shafarevich group'', $\Sha(J(Y))$, of the Jacobian fibration $J(Y)$ of Calabi--Yau manifold $Y$. The discrete part of the Tate--Shafarevich group is then identified with the discrete gauge group that forms in F-theory compactification on $Y$ \cite{BDHKMMS}. The set of Calabi--Yau genus-one fibrations, whose Jacobians are isomorphic to the specific Jacobian fibration $J$, is known to have a group structure, and this group yields the Tate--Shafarevich group $\Sha(J)$ of the Jacobian fibration $J$.} with a multisection of degree $n$, or an ``$n$-section.'' Here, we study F-theory models with a discrete $\Z_4$ gauge symmetry on genus-one fibrations over general base spaces, as well as F-theory models with a discrete $\Z_5$ gauge symmetry on genus-one fibrations over ``Fano-type'' base spaces. F-theory models with discrete $\Z_2$ \cite{MTsection} and $\Z_3$ \cite{KMOPR, CDKPP} gauge symmetries have been systematically analyzed on genus-one fibrations over general base spaces. In addition, F-theory models with a discrete $\Z_4$ gauge symmetry were studied in \cite{BGKintfiber, Kdisc}, and four-dimensional (4D) F-theory models with a discrete $\Z_5$ symmetry were constructed in \cite{Kdisc}.
\par We construct 6D F-theory models with a discrete gauge group, and in doing so we utilize a special class of ``Fano 3-folds,'' known in mathematics as ``del Pezzo 3-folds.'' Fano 3-folds are a generalization of projective spaces. Furthermore, del Pezzo 3-folds are a class of Fano 3-folds that have a notion of ``degrees.'' Genus-one fibered Calabi--Yau 3-folds \footnote{Structures of elliptic fibrations of 3-folds are discussed in \cite{Nak, DG, G}.} without a section can be constructed using these del Pezzo 3-folds, and discrete gauge groups of different degrees arise in F-theory on genus-one fibered Calabi--Yau 3-folds constructed from del Pezzo 3-folds, depending on the degrees of the del Pezzo 3-folds. We consequently obtain 6D F-theory models with discrete $\Z_5$, $\Z_4$, $\Z_3$, and $\Z_2$ gauge groups. Specifically, we obtain 6D models with discrete $\Z_5$ and $\Z_4$ gauge symmetries. As explained in Section \ref{ssec3.1}, 6D F-theory models on genus-one fibered Calabi--Yau 3-folds are obtained over any base space for a discrete $\Z_4$ gauge group. Six-dimensional F-theory models on genus-one fibered Calabi--Yau 3-folds with $\P^1\times\P^1$, $\P^2$, and del Pezzo surfaces as base spaces are obtained with a discrete $\Z_5$ gauge group. We first construct genus-one fibered Calabi--Yau 3-folds over the bases $\P^1\times\P^1$ and $\P^2$, and then we find that this construction in fact generalizes to Calabi--Yau genus-one fibrations over any base space for a discrete $\Z_4$ gauge group. For models with a discrete $\Z_5$ gauge group, by applying the Riemann--Roch theorem, we observe that the construction of genus-one fibered Calabi--Yau 3-folds with a multisection of degree five is possible over del Pezzo surfaces. In \cite{Kdisc}, genus-one fibered Calabi--Yau 4-folds lacking a global section were constructed using del Pezzo 3-folds, and 4D F-theory models with discrete $\Z_5$, $\Z_4$, $\Z_3$, and $\Z_2$ gauge groups were obtained on these Calabi--Yau 4-folds. The discussion in \cite{Kdisc} mainly focused on the constructions of genus-one fibered Calabi--Yau 4-folds with bases consisting of del Pezzo 3-folds. As discussed in Section \ref{ssec3.2}, we find that for discrete $\Z_4, \Z_3,$ and $\Z_2$ gauge groups the results in \cite{Kdisc} can be extended to genus-one fibered Calabi--Yau 4-folds over any base. By applying the Riemann--Roch theorem, we show that the models with a discrete $\Z_5$ gauge group in \cite{Kdisc} extend to a wider class of genus-one fibered Calabi--Yau 4-folds over more general bases of Fano 3-folds. 

\par Local model buildings \cite{DWmodel, BHV1, BHV2, DW} of F-theory compactifications have been emphasized in recent studies. Global structures of F-theory models, however, need to be analyzed to address the issues of gravity and the early universe including inflation. We study the compactification geometries from the global perspective in this note.

\vspace{0.5cm}

\par After outlining the main results of this work in Section \ref{sec2}, we construct families of genus-one fibered Calabi--Yau 3-folds without a section by utilizing del Pezzo 3-folds in Section \ref{sec3}. The resulting Calabi--Yau 3-folds have multisections of degrees two to five, namely a bisection, 3-section, 4-section, and 5-section, according to the degrees of the del Pezzo 3-folds. Discrete gauge groups arise in F-theory compactifications on the constructed Calabi--Yau 3-folds with degrees corresponding to those of the multisections. We first discuss constructions of Calabi--Yau genus-one fibrations over the base $\P^1\times\P^1$ and $\P^2$, and then we find that this construction generalizes to Calabi--Yau genus-one fibrations over general base surfaces for models with discrete $\Z_4$, $\Z_3$, and $\Z_2$ gauge groups. For models with a discrete $\Z_5$ gauge group, we find that this construction extends to genus-one fibered Calabi--Yau 3-folds over del Pezzo surfaces.
\par In Section \ref{sec4}, we consider F-theory compactifications on the Calabi--Yau genus-one fibrations constructed in Section \ref{sec3}, yielding 6D F-theory with discrete gauge symmetries. We obtain 6D F-theory models with discrete $\Z_5$, $\Z_4$, $\Z_3$, and $\Z_2$ gauge groups. Specifically, these include models on Calabi--Yau genus-one fibrations (over general base surfaces) with a discrete $\Z_4$ gauge group. For a discrete $\Z_5$ gauge group, we obtain 6D F-theory models on Calabi--Yau genus-one fibrations over del Pezzo surfaces, $\P^1\times\P^1$, and $\P^2$. In addition, we argue that an analogue of a discussion in Section \ref{ssec3.1} extends the results obtained in \cite{Kdisc} to a wider class of genus-one fibered Calabi--Yau 4-folds over general base 3-folds for 4D F-theory models with discrete $\Z_4$, $\Z_3$, and $\Z_2$ gauge groups. For models with a discrete $\Z_5$ gauge group, we find that the result in \cite{Kdisc} extends to 4D F-theory models over more general Fano 3-folds as bases. 
\par We also discuss a process, where a discrete $\Z_2$ gauge group transitions to a discrete $\Z_4$ gauge group. This effect can be viewed as the reverse of the process where a 4-section splits into a pair of bisections when the defining equation of a genus-one fibration takes a special form in the moduli of 4-section geometries. This process appears somewhat puzzling from a physical viewpoint, because it seems natural to expect that a discrete gauge group breaks down to another discrete gauge group of a smaller degree, as we will discuss in section \ref{ssec4.2}. We present concluding remarks in Section \ref{sec5}.

\section{Summary of the results}
\label{sec2}

\par A discrete $\Z_n$ gauge group arises in F-theory on a genus-one fibration with an $n$-section. There are various approaches to constructing genus-one fibrations lacking a global section. However, regarding the classification and understanding of discrete gauge groups from a physical viewpoint, the construction of genus-one fibrations with a multisection of a specific degree is an essential problem. Fano 3-folds are a generalization of projective 3-folds, and the special class of Fano 3-folds called del Pezzo 3-folds \footnote{This special class of Fano 3-folds is also referred to as {\it Fano 3-folds of index two}, as in \cite{Isk}. We refer to them as del Pezzo 3-folds in this note, following \cite{Fujita1, Fujita2}.} has a very useful property for constructing genus-one fibered Calabi--Yau manifolds with multisections of specific degrees and without a section. The structure of del Pezzo 3-folds $V_n$ of degree $n$ is described in Section \ref{ssec3.1} \footnote{We do not consider the del Pezzo 3-fold of degree 1, $V_1$ \cite{Fujita3} in this study.}. The complete intersections of two hypersurfaces in the products $V_n\times \P^1\times\P^1$ and $V_n\times \P^2$ yield genus-one fibered Calabi--Yau 3-folds. This is a consequence of the property of del Pezzo 3-folds that their canonical class $K$ is isomorphic to ${\cal O}_{V_n}(-2)$. Using an argument similar to that presented in \cite{Kdisc}, one finds that the resulting genus-one fibered Calabi--Yau 3-folds do not have a global section to the fibration. 
\par The constructed Calabi--Yau 3-folds $Y$ naturally yield projections onto $\P^1\times \P^1$ (or $\P^2$, respectively) and $V_n$. 

$$
\begin{CD}
Y @>{q}>> V_n \\
@V{p}VV \\
\P^1\times \P^1\hspace{2mm} ({\rm or}\hspace{2mm} \P^2).
\end{CD}
$$ 

$\P^1\times\P^1$ (or $\P^2$, respectively) is regarded as the base of the genus-one fibration, and the pullback of a hyperplane class ${\cal O}(1)$ in $V_n$ yields an $n$-section for del Pezzo 3-folds $V_n$ of degrees $n=2,3,4,5$ \cite{Kdisc}. F-theory compactifications on these genus-one fibered Calabi--Yau 3-folds yield 6D theories with discrete $\Z_5$, $\Z_4$, $\Z_3$, and $\Z_2$ gauge groups over the base $\P^1\times\P^1$ and $\P^2$. When $n=6$, one has three bisections, and the case of $n=8$ with a del Pezzo 3-fold $V_8$ yields a 4-section \cite{Kdisc}. When $n=7$, the Calabi--Yau 3-fold constructed from $V_7$ admits a global section \cite{Kdisc}. We do not consider the del Pezzo 3-fold $V_7$ of degree seven. 
\par The constructions of genus-one fibered Calabi--Yau 3-folds over $\P^1\times\P^1$ and $\P^2$ in fact generalize to Calabi--Yau genus-one fibrations over general base surfaces for del Pezzo 3-folds $V_n$ of degrees $n=2,3,4,6,8$. These yield Calabi--Yau genus-one fibrations with multisections of degrees between two and four over general base surfaces. Therefore, F-theory compactifications on these Calabi--Yau 3-folds yield 6D theories with discrete $\Z_4, \Z_3,$ and $\Z_2$ gauge groups over any base surfaces. The construction of Calabi--Yau 3-folds over $\P^1\times\P^1$ and $\P^2$ means that the coefficients of the equation of the genus-one fibration are sections of line bundles over $\P^1\times\P^1$ and $\P^2$. We now want to generalize the construction of a Calabi--Yau genus-one fibration over $\P^1\times\P^1$ (and $\P^2$) to a general base space $B$, using an equation of the same form for the genus-one fibration. To this end, one requires that the total space of the genus-one fibration over the base $B$ yields a Calabi--Yau 3-fold. This condition imposes constraints on the coefficients of the equation such that they are sections of line bundles corresponding to specific divisor classes, so that when the Jacobian fibration $y^2=x^3+f\, x+g$ is taken the Weierstrass coefficients of the Jacobian fibration, $f$ and $g$, satisfy the relations $[f]=-4K$ and $[g]=-6K$, namely, $f$ and $g$ are sections of the line bundles corresponding to the divisor classes $-4K$ and $-6K$. Here, $K$ denotes the canonical class of the base $B$. Therefore, when appropriate line bundles are chosen they have sections over $B$, and the genus-one fibered Calabi--Yau construction over $\P^1\times\P^1$ (and $\P^2$) generalizes to one over the base $B$.
\par From the aforementioned argument, we deduce that the Calabi--Yau 3-fold construction with multisections of degrees between two and four over $\P^1\times\P^1$ (and $\P^2$) generalizes to the construction of genus-one fibered Calabi--Yau 3-folds over general base surfaces. 
\par From these, we learn that there are constructions of genus-one fibered Calabi--Yau 3-folds lacking a global section with multisections of degrees between two and four over any base surfaces. 
\par F-theory compactifications on these genus-one fibered Calabi--Yau 3-folds yield 6D models with discrete $\Z_4$, $\Z_3$, and $\Z_2$ gauge groups over any base surfaces. 
\par We find that the construction of genus-one fibered Calabi--Yau 3-folds with a 5-section using the del Pezzo 3-fold $V_5$ extends to Calabi--Yau 3-folds over del Pezzo surfaces as base surfaces. Therefore, we obtain F-theory models with a discrete $\Z_5$ gauge group over del Pezzo surfaces, $\P^1\times\P^1$, and $\P^2$. Computing the explicit Jacobian fibration of the Calabi--Yau genus-one fibration with a 5-section constructed in this study is considerably difficult, and this presents an obstacle to generalizing the construction to Calabi--Yau genus-one fibrations over any base.
\par F-theory compactifications on genus-one fibered Calabi--Yau manifolds over general base spaces were analyzed for a discrete $\Z_2$ and $\Z_3$ gauge groups in \cite{MTsection} and \cite{KMOPR, CDKPP}, respectively. We specifically discuss F-theory compactifications on genus-one fibered Calabi--Yau manifolds over general base spaces with discrete $\Z_4$ and $\Z_5$ gauge groups in Section \ref{ssec3.1}. Although our argument in Section \ref{ssec3.1} focuses on 6D theories, our argument does not require a specific dimension, and the constructions of genus-one fibered Calabi--Yau manifolds with multisections apply to other dimensions. In particular, these apply to 4D theories.

\section{Genus-one fibered Calabi--Yau 3-folds with multisections of degrees two to five}
\label{sec3}

\subsection{Constructions of Calabi--Yau 3-folds with multisections}
\label{ssec3.1}
We first describe constructions of genus-one fibered Calabi--Yau 3-folds without a global section over the bases $\P^1\times\P^1$ and $\P^2$, for simplicity. As will be explained after these constructions, similar Calabi--Yau constructions are in fact possible for genus-one fibrations over general base surfaces (except Calabi--Yau constructions with a 5-section: as we will see, the Calabi--Yau constructions with a 5-section extend to Calabi--Yau genus-one fibrations over del Pezzo surfaces). 
\par Del Pezzo 3-folds are Fano 3-folds, for which the canonical bundles $K$ are isomorphic to ${\cal O}(-2)$ \cite{Fujita1}. This is a useful property for our purposes, and implies that two complete intersections of hypersurfaces in del Pezzo 3-folds times products of projective spaces yield genus-one fibered Calabi--Yau manifolds. Del Pezzo 3-folds are classified according to their degrees \cite{Isk, Fujita1, Fujita2}. The results of classifying del Pezzo 3-folds are presented in Table \ref{tablistofdelpezzo3-folds}. (The results in Table \ref{tablistofdelpezzo3-folds} are given in \cite{Isk, Fujita1, Fujita2}.) 

\begingroup
\renewcommand{\arraystretch}{1.5}
\begin{table}[htb]
\begin{center}
  \begin{tabular}{|c|c|} \hline
 Degree $n$ of del Pezzo 3-fold $V_n$ & Structure  \\ \hline
 2 & Double cover of $\P^3$  \\
 3 &  Cubic 3-fold in $\P^4$ \\ 
 4 &  Complete intersection of two quadric hypersurfaces in $\P^5$ \\
 5 &  Intersection of three hyperplane sections in $Gr(2,5)\subset \P^9$ \\ 
 6 &  $\P^1\times\P^1\times\P^1$ \\
 7 &  1 point Blow-up of $\P^3$ \\
 8 &  $\P^3$ \\ \hline   
\end{tabular}
\caption{\label{tablistofdelpezzo3-folds}Del Pezzo 3-folds of degree $d=2, \cdots, 8$ \cite{Isk, Fujita1, Fujita2} are listed. $Gr(2,5)$ in the table represents the complex Grassmannian of two-dimensional linear subspaces in $\C^5$. $Gr(2,5)$ is embedded inside $\P^9$ via the Pl\"ucker embedding, and the del Pezzo 3-fold $V_8$ of degree $8$ is the image of $\P^3$ embedded in $\P^9$ via the Veronese embedding.}
\end{center}
\end{table}  
\endgroup 

\par The del Pezzo 3-fold of degree $n$ is denoted by $V_n$. The complete intersection of two hyperplane classes ${\cal O}_{V_n\times\P^1\times\P^1}(1,1,1)$ in the product
\begin{equation}
V_n \times \P^1 \times \P^1
\end{equation}
yields a genus-one fibered Calabi--Yau 3-fold $Y$. By construction, the Calabi--Yau 3-fold $Y$ naturally yields projections onto $\P^1\times\P^1$ and $V_n$. 
$$
\begin{CD}
Y @>{q}>> V_n \\
@V{p}VV \\
\P^1\times \P^1
\end{CD}
$$ 
The projection onto $\P^1\times\P^1$ yields a genus-one fibration. By an argument similar to that in \cite{Kdisc}, this genus-one fibration does not have a rational section. Furthermore, the projection onto $\P^1$ yields a K3 fibration. 
\par If we denote the projection onto $V_n$ by $q$, then the pullback of a hyperplane class ${\cal O}_{V_n}(1)$ in $V_n$, 
\begin{equation}
\label{pullback in 3.1}
q^*{\cal O}_{V_n}(1)
\end{equation}
gives an $n$-section for degrees $n=2,3,4,5$ \cite{Kdisc}. For $n=6$, the pullback (\ref{pullback in 3.1}) splits into three bisections, and for $n=8$ the pullback (\ref{pullback in 3.1}) yields a 4-section \cite{Kdisc}. 
\par An argument similar to that given above applies to the construction of genus-one fibered Calabi--Yau 3-folds over the base $\P^2$. The complete intersection of ${\cal O}_{V_n\times \P^2}(1,1)$ and ${\cal O}_{V_n\times \P^2}(1,2)$ yields a genus-one fibered Calabi--Yau 3-fold, and this Calabi--Yau manifold does not have a global section. The Calabi--Yau 3-fold naturally yields projections onto $\P^2$ and $V_n$.
$$
\begin{CD}
Y @>{q}>> V_n \\
@V{p}VV \\
\P^2
\end{CD}
$$ 
The projection onto $\P^2$ gives a genus-one fibration over $\P^2$, and the pullback of the hyperplane class ${\cal O}_{V_n}(1)$ under the projection $q$ onto $V_n$, 
\begin{equation}
q^*{\cal O}_{V_n}(1)
\end{equation}
gives an $n$-section for $n=2,3,4,5$. Here $n=6$ yields three bisections, and $n=8$ gives a 4-section, similarly to what we stated for the constructions of genus-one fibered Calabi--Yau 3-folds over $\P^1\times\P^1$.

\vspace{1cm}

\par We have focused on the constructions of Calabi--Yau genus-one fibrations over the specific bases $\P^1\times \P^1$ and $\P^2$. However, these constructions in fact generalize to genus-one fibered Calabi--Yau 3-folds over any base surface, except for the construction of a Calabi--Yau 3-fold with a 5-section using the del Pezzo 3-fold $V_5$. In other words, it is possible to construct families of genus-one fibered Calabi--Yau 3-folds lacking a global section that possess $n$-sections for $n=2,3,4$ over any base surface. Let us provide a sketch of a proof of this, taking genus-one fibered Calabi--Yau 3-folds possessing a 4-section as an example. The construction of Calabi--Yau 3-folds with a 5-section over $\P^1\times\P^1$ and $\P^2$ generalizes to Calabi--Yau 3-folds over del Pezzo surfaces. We will also discuss this. Calabi--Yau genus-one fibrations with a bisection \cite{BM, MTsection} and with a 3-section \cite{KMOPR, CDKPP} over general base spaces have previously been analyzed. Analyzing Calabi--Yau genus-one fibration constructions with a 4-section over general base surfaces and Calabi--Yau genus-one fibration constructions with a 5-section over the del Pezzo surfaces can be interesting, because these analyses directly relate to discrete $\Z_4$ and $\Z_5$ gauge groups in string compactifications. 
\par Using the del Pezzo 3-folds $V_4$ and $V_8$, we constructed Calabi--Yau 3-folds with a 4-section over $\P^1\times\P^1$ and $\P^2$. The genus-one fiber of the construction involving $V_8$ is the complete intersection of two quadric hypersurfaces in $\P^3$ \footnote{Because $V_8$ is $\P^3$ embedded in $\P^9$ via the Veronese embedding, complete intersection of two hyperplanes classes ${\cal O}_{V_8}(1)$ in $V_8$ is equivalent to complete intersection of two quadric hypersurfaces in $\P^3$ in the usual sense.} \footnote{A method to compute the Jacobian fibration of a genus-one fibration constructed as a complete intersection is also discussed in \cite{BGKintfiber}. We take a different approach to compute the Jacobian fibration of a genus-one fibration constructed as a complete intersection in this study.}. We choose this construction, and generalize it to the Calabi--Yau genus-one fibration with a 4-section over any base surface. Using a mathematical technique, the Jacobian of this type of genus-one fibration can be explicitly constructed over any base space, as we will show. Imposing some conditions on the Weierstrass coefficients of the resulting Jacobian ensures that this indeed yields a Calabi--Yau Jacobian fibration. Then, this implies that the original genus-one fibration is Calabi--Yau. 
\par The complete intersection can be written as
\begin{eqnarray}
\label{complete intersection in 3.1}
\begin{split}
F_1= & a_1\, x_1^2+a_2\, x_2^2+a_3\, x_3^2+a_4\, x_4^2+2a_5\, x_1x_2 \\
& +2a_6\, x_1x_3+ 2a_7\, x_1x_4+2a_8\, x_2x_3+ 2a_9\, x_2x_4+2a_{10}\, x_3x_4=0 \\ 
F_2= & b_1\, x_1^2+b_2\, x_2^2+b_3\, x_3^2+b_4\, x_4^2+2b_5\, x_1x_2 \\
& +2b_6\, x_1x_3+ 2b_7\, x_1x_4+2b_8\, x_2x_3+ 2b_9\, x_2x_4+2b_{10}\, x_3x_4=0,
\end{split}
\end{eqnarray}
where we have used $[x_1:x_2:x_3:x_4]$ to denote the coordinates of $\P^3$, and $a_i$ and $b_j$ for $i,j=1, \cdots, 10$ are sections of line bundles of the base space $B$. The factors of two before $a_5, \cdots a_{10}$ and $b_5, \cdots b_{10}$ are a convention. We introduce a variable $\lambda$, and subtract $\lambda$ times $F_2$ from $F_1$. The coefficients of $F_1-\lambda\, F_2$ are arranged into a 4 $\times$ 4 symmetric matrix:
\begin{equation}
\label{symmetricmatrix in 3.1}
\begin{pmatrix}
a_1-\lambda \, b_1 & a_5-\lambda\, b_5 & a_6-\lambda \, b_6 & a_7-\lambda \, b_7 \\
a_5-\lambda\, b_5 & a_2-\lambda \, b_2 & a_8-\lambda \, b_8 & a_9-\lambda \, b_9 \\
a_6-\lambda \, b_6 & a_8-\lambda\, b_8 & a_3-\lambda \, b_3 & a_{10}-\lambda \, b_{10} \\
a_7-\lambda \, b_7 & a_9-\lambda\, b_9 & a_{10}-\lambda \, b_{10} & a_4-\lambda \, b_4
\end{pmatrix}.
\end{equation}
The determinant of the resulting symmetric matrix (\ref{symmetricmatrix in 3.1}) yields a quartic polynomial in $\lambda$, and equating this polynomial with $\tau^2$ gives a double cover of the quartic polynomial:
\begin{equation}
\label{quartic polynomial in 3.1}
\begin{split}
\tau^2 & = {\rm det}\, \begin{pmatrix}
a_1-\lambda \, b_1 & a_5-\lambda\, b_5 & a_6-\lambda \, b_6 & a_7-\lambda \, b_7 \\
a_5-\lambda\, b_5 & a_2-\lambda \, b_2 & a_8-\lambda \, b_8 & a_9-\lambda \, b_9 \\
a_6-\lambda \, b_6 & a_8-\lambda\, b_8 & a_3-\lambda \, b_3 & a_{10}-\lambda \, b_{10} \\
a_7-\lambda \, b_7 & a_9-\lambda\, b_9 & a_{10}-\lambda \, b_{10} & a_4-\lambda \, b_4
\end{pmatrix} \\
& = \alpha\, \lambda^4 + \beta\, \lambda^3 + \gamma\, \lambda^2 + \delta\, \lambda + \omega.
\end{split}
\end{equation}
where $\alpha, \beta, \gamma, \delta, \omega$ are polynomials of $a_i$ and $b_j$. The Jacobian fibration of the resulting double cover of the quartic polynomial (\ref{quartic polynomial in 3.1}) yields the Jacobian fibration \cite{BM} of the complete intersection (\ref{complete intersection in 3.1}). A discussion of the construction of the Jacobian fibration from the double cover of the quartic polynomial is given in \cite{BM, MTsection}. In short, constructing the Jacobian of the complete intersection (\ref{complete intersection in 3.1}) involves two steps. First, one computes the double cover of the quartic polynomial from the coefficients of the complete intersection. Then, one takes the Jacobian fibration of the obtained double cover of the quartic polynomial to deduce the Jacobian fibration of the complete intersection (\ref{complete intersection in 3.1}). The coefficients $f,g$ of the resulting Weierstrass equation $y^2=x^3+f\, x+g$ of the Jacobian fibration are polynomials in the coefficients $a_i$ and $b_j$ of the complete intersection (\ref{complete intersection in 3.1}). 
\par The condition that the Jacobian of the complete intersection (\ref{complete intersection in 3.1}) describes a Calabi--Yau 3-fold requires that 
\begin{eqnarray}
\label{CY condition Weierstrass in 3.1}
[f] & = & -4K \\ \nonumber
[g] & = & -6K,
\end{eqnarray}
where $K$ denotes the canonical class of the base space $B$. When this is satisfied, the complete intersection (\ref{complete intersection in 3.1}) is also Calabi--Yau. The condition (\ref{CY condition Weierstrass in 3.1}) is satisfied when the divisors corresponding to the coefficients $\alpha, \beta, \gamma, \delta, \omega$ satisfy the following relations \cite{MTsection}:
\begin{eqnarray}
\label{relations quartic in 3.1}
[\alpha] & = & 2L \\ \nonumber
[\beta] & = & -K+L \\ \nonumber 
[\gamma] & = & -2K \\ \nonumber
[\delta] & = & -3K-L \\ \nonumber
[\omega] & = & -4K-2L,
\end{eqnarray}
where $L$ is a line bundle selected such that $[\beta]$ and $[\delta]$ are effective \cite{MTsection}, and $\alpha, \beta, \gamma, \delta, \omega$ are polynomials in $a_i$ and $b_j$ of the complete intersection (\ref{complete intersection in 3.1}). Therefore, the relations (\ref{relations quartic in 3.1}) impose constraints on the coefficients $a_i$ and $b_j$. When one chooses appropriate line bundles over the base $B$ such the coefficients $a_i$ and $b_j$ satisfy (\ref{relations quartic in 3.1}), the complete intersection (\ref{complete intersection in 3.1}) yields a Calabi--Yau manifold. This describes the construction of a genus-one fibered Calabi--Yau manifold with a 4-section that lacks a global section, constructed as in (\ref{complete intersection in 3.1}) over any base $B$. Because this argument does not depend on the dimensionality of the base $B$, this construction yields a Calabi--Yau manifold of any dimension. In particular, this construction also yields genus-one fibered Calabi--Yau 4-folds with a 4-section that lack a global section over any base 3-fold, as well as genus-one fibered Calabi--Yau 3-folds with a 4-section over any base surface. 

\vspace{1cm}

\par We would like to extend the construction of Calabi--Yau 3-folds with a 5-section over $\P^1\times\P^1$ (or $\P^2$) using the del Pezzo 3-fold $V_5$ discussed previously to Calabi--Yau genus-one fibrations over other base surfaces. To generalize the construction to other base spaces $B$, one needs to consider the fibering of intersection of five hyperplane classes ${\cal O}(1)$ in the Grassmannian $Gr(2,5)$ embedded in $\P^9$ as a genus-one fiber over the base $B$. A possible obstruction to this generalization is that computing the explicit Weierstrass equation of the Jacobian fibration of this genus-one fibration is considerably challenging. Instead of attempting to generalize to fibrations over any base space, we show that the construction of models with a discrete $\Z_5$ gauge group over the base $\P^1\times\P^1$ and $\P^2$ extends to models over the base del Pezzo surfaces. One can generalize the previous construction to genus-one fibrations over del Pezzo surfaces owing to a useful property of del Pezzo surfaces, without computing the Weierstrass equation of the Jacobian fibration. By an argument similar to that in \cite{MTsection}, requiring that the Weierstrass coefficients $f,g$ of the Jacobian fibration $y^2=x^3+f\, x+g$ are sections of the line bundles corresponding to the divisor classes $-4K$ and $-6K$ implies that the coefficients of the equation of the genus-one fibration are sections of line bundles that correspond to positive integer multiples of $-K$, (integer multiples of) some line bundle $L$, or sums of positive integer multiples of $-K$ and integer multiples of $L$. One has a Calabi--Yau genus-one fibration over a base $B$ if these line bundles have sections over the base $B$. When the anticanonical class $-K$ is ample, the line bundles indeed have sections over the base $B$ if one selects the line bundle $L$ as some appropriate positive integer multiple of the anticanonical class $-K$, as a consequence of the Riemann--Roch theorem. Algebraic surfaces with ample anticanonical classes are precisely the del Pezzo surfaces, and therefore we have learned that the construction of Calabi--Yau genus-one fibrations with a 5-section over $\P^1\times\P^1$ and $\P^2$ actually extends to Calabi--Yau genus-one fibrations with a 5-section over del Pezzo surfaces. The argument that we have employed does not depend on the dimensionality of the base space $B$. Our argument also applies to the construction of genus-one fibered Calabi--Yau 4-folds. 

\vspace{1cm}

\par As will be discussed in Section \ref{sec4}, F-theory compactifications on the resulting Calabi--Yau 3-folds yield 6D models with discrete $\Z_4$, $\Z_3$, and $\Z_2$ gauge groups over any base surface, and 6D models with a discrete $\Z_5$ gauge group over del Pezzo surfaces, $\P^1\times\P^1$, and $\P^2$. 

\subsection{Application to Calabi--Yau 4-folds}
\label{ssec3.2}
\par The constructions of genus-one fibered Calabi--Yau 4-folds in \cite{Kdisc} focused on genus-one fibrations over del Pezzo 3-folds $V_n$ (including $\P^3$ and $\P^1\times\P^1\times\P^1$) as base. We found that Calabi--Yau 3-fold constructions as genus-one fibrations over $\P^1\times\P^1$ and $\P^2$ generalize to genus-one fibrations over any base surface for $n=4,8$ in Section \ref{ssec3.1}, and similarly for $n=2,3,6$. For Calabi--Yau 3-folds using the del Pezzo 3-fold $V_5$, the construction generalizes to Calabi--Yau 3-folds with a 5-section over del Pezzo surfaces. These arguments do not depend on the dimensionality, and thus they also apply to genus-one fibered Calabi--Yau 4-folds. The constructions of genus-one fibered Calabi--Yau 4-folds utilizing del Pezzo 3-folds in \cite{Kdisc} therefore generalize to Calabi--Yau genus-one fibrations over any base 3-folds for constructions using del Pezzo 3-folds $V_n$ of degrees $n=2,3,4,6,8$. From these, we find that the constructions of Calabi--Yau 4-folds in \cite{Kdisc} generalize to genus-one fibered Calabi--Yau 4-folds without a section with multisections of degrees between two and four over any base 3-folds. For the genus-one fibered Calabi--Yau 4-folds with a 5-section constructed in \cite{Kdisc}, because 3-folds with ample anticanonical classes are precisely the Fano 3-folds, the constructions of Calabi--Yau 4-folds over del Pezzo 3-folds extend to Calabi--Yau 4-folds over Fano 3-folds. 
\par We have learned that the 4D F-theory models with discrete $\Z_4$, $\Z_3$, and $\Z_2$ gauge groups over the base del Pezzo 3-folds obtained in \cite{Kdisc} extend to 4D F-theory models with discrete $\Z_4$, $\Z_3$, and $\Z_2$ gauge groups over any base 3-folds. The 4D F-theory models with a discrete $\Z_5$ gauge group over the base del Pezzo 3-folds obtained in \cite{Kdisc} extend to 4D F-theory models with a discrete $\Z_5$ gauge group over any Fano 3-folds.

\section{Discrete $\Z_5, \Z_4, \Z_3,$ and $\Z_2$ gauge groups in six-dimensional F-theory models}
\label{sec4}

\subsection{Discrete $\Z_5, \Z_4, \Z_3,$ and $\Z_2$ gauge groups}
\label{ssec4.1}
\par We obtain 6D F-theory models with discrete gauge symmetries when we consider compactifications on the genus-one fibered Calabi--Yau 3-folds constructed in Section \ref{ssec3.1}. As we saw, the Calabi--Yau 3-folds have $n$-sections with $n=2,3,4,5$, although they lack a global section. Discrete $\Z_5, \Z_4, \Z_3,$ and $\Z_2$ gauge groups arise in 6D F-theory models on the constructed Calabi--Yau 3-folds, depending on the degrees of the multisections. Because these Calabi--Yau manifolds are constructed as genus-one fibrations over any base surfaces for $n=2,3,4$, we obtain 6D F-theory models with discrete gauge groups over any base surfaces. These include models with a discrete $\Z_4$ gauge group. For Calabi--Yau 3-folds with a 5-section, we obtained F-theory models with a discrete $\Z_5$ gauge group over any del Pezzo surface (as well as over $\P^1\times\P^1$ and $\P^2$) as base.
\par As we saw in Section \ref{ssec3.2}, the argument given in Section \ref{ssec3.1} applies to the 4D F-theory models with discrete gauge groups constructed in \cite{Kdisc}. The compactification geometries of 4D F-theory models with discrete gauge groups discussed in \cite{Kdisc} are genus-one fibrations, mainly over base del Pezzo 3-folds. Our results generalize the constructions in \cite{Kdisc} to 4D F-theory models with discrete $\Z_4, \Z_3,$ and $\Z_2$ gauge groups on Calabi--Yau genus-one fibrations over any base 3-folds. Specifically, we have obtained 4D F-theory models with a discrete $\Z_4$ gauge group over any base space. Our results also generalize the 4D F-theory models with a discrete $\Z_5$ gauge group in \cite{Kdisc} to models with a discrete $\Z_5$ gauge group over any Fano 3-folds as base, which are more general than del Pezzo 3-folds. 

\subsection{Transition in discrete $\Z_4$ gauge group}
\label{ssec4.2}
A discrete $\Z_4$ gauge group arises in F-theory on genus-one fibered Calabi--Yau 3-folds with a 4-section constructed using the del Pezzo 3-folds $V_4$ and $V_8$. In the moduli of these Calabi--Yau 3-folds with a 4-section, when the defining equation of the Calabi--Yau manifold takes a special form the 4-section splits into a pair of bisections. The discrete gauge group formed in F-theory compactification in this situation is $\Z_2$. A phenomenon of this kind was previously observed for 4D F-theory in \cite{Kdisc}. 
\par For simplicity, we focus on the construction using $V_8$. The general complete intersection of two quadric hypersurfaces in $\P^3$ fibered over a base surface admits a 4-section, and a discrete $\Z_4$ gauge group forms in the F-theory compactification, as analyzed in \cite{BGKintfiber, Kdisc}. When we consider complete intersections of the form
\begin{eqnarray}
\label{complete intersection form 1 in 4.2}
x_1^2 + x_3^2 +2f\, x_2x_4 = 0 \\ \nonumber
x_2^2 + x_4^2 +2g\, x_1x_3 = 0
\end{eqnarray}
where $[x_1:x_2:x_3:x_4]$ denotes the coordinates on $\P^3$ and $f,g$ are sections of line bundles over the base $B$, the 4-section splits into a pair of bisections, by an argument similar to that in \cite{Kdisc}. $\{x_1=0, \hspace{1mm} x_2=i x_4\}$ and $\{x_1=0, \hspace{1mm} x_2=-i x_4\}$ yield bisections \cite{Kdisc}. 
\par We would like to point out that the splitting of a 4-section into a pair of bisections in fact occurs in complete intersections that are more general than the form (\ref{complete intersection form 1 in 4.2}) considered in \cite{Kdisc}. 4-section splits into bisections in complete intersections given in the following form:
\begin{eqnarray}
\label{complete intersection form 2 in 4.2}
a_1\, x_1^2 +a_2\, x_2^2 + x_3^2 + a_4\, x_4^2+ 2a_5\, x_1x_2+2a_6\, x_1x_3+2a_7\, x_1x_4 +2a_9\, x_2x_4 = 0 \\ \nonumber
b_2\, (x_2^2 + x_4^2) +b_1\, x_1^2 +2b_5\, x_1x_2+ 2b_6\, x_1x_3 +2b_7\, x_1x_4 = 0,
\end{eqnarray}
where $a_i$ and $b_j$ are sections of line bundles over the base $B$. $\{x_1=0, \hspace{1mm} x_2=i x_4\}$ and $\{x_1=0, \hspace{1mm} x_2=-i x_4\}$ still yield bisections to the complete intersection (\ref{complete intersection form 2 in 4.2}). 
\par The splitting of 4-section into a pair of bisections can be viewed as a process where a model transitions from the 4-section geometry of genus-one fibered Calabi--Yau 3-folds to a bisection geometry. This can also be viewed as the reverse of the transition from a discrete $\Z_2$ gauge group to a discrete $\Z_4$ gauge group, along the lines of the argument given in \cite{MTsection}. 
\par This phenomenon appears unnatural from a physical viewpoint. Let us explain this point. The process in which a 4-section splits into a pair of bisections can be seen as an intermediate step of a 4-section splitting into four sheets of global sections. Taking the limit at which a 4-section splits into four global sections in the moduli of 4-section geometries can be seen as the reverse of the process in which a $U(1)^3$ gauge symmetry breaks to a discrete $\Z_4$ gauge group as discussed in \cite{MTsection}. Because a 4-section splitting into a pair of bisections is an intermediate of the process where a 4-seciton splits into four sheets of global sections, one naturally expects that the reverse of the process, where a 4-section splits into a pair of bisections, and the bisections further split into global sections, corresponds to a $U(1)$ gauge symmetry breaking down to some discrete gauge group, and this intermediate discrete gauge group further breaking down to a discrete $\Z_4$ gauge group. However, we observed that a discrete $\Z_2$ gauge group transitions to a discrete $\Z_4$ gauge group in the examples that we have just discussed. This appears a discrete gauge group enhanced to another discrete gauge group of a higher degree, rather than a discrete gauge group breaking down to another discrete gauge group of a smaller degree.
\par Because the other discrete gauge groups discussed in this note, $\Z_2$, $\Z_3$, and $\Z_5$, have prime degrees, when the corresponding multisections split into multisections of smaller degrees they should contain a global section. Therefore, we deduce that when this occurs models with discrete $\Z_2$, $\Z_3$, and $\Z_5$ gauge groups become models in which a discrete gauge group does not arise. The transition from a discrete gauge group to a discrete $\Z_2$, $\Z_3$, or $\Z_5$ gauge group, in the way similar to that of a discrete $\Z_2$ gauge group to a discrete $\Z_4$ gauge group that we have just discussed, does not occur in the moduli of multisection geometries.

\section{Open problems and conclusions}
\label{sec5}
\par We constructed 6D F-theory models with various discrete gauge groups. These include F-theory models with discrete $\Z_4$ gauge groups on genus-one fibrations over any base surface, and with a discrete $\Z_5$ gauge group on genus-one fibrations over del Pezzo surfaces, $\P^1\times\P^1$, and $\P^2$. 
\par The discussion in this study applies to 4D F-theory models with discrete gauge groups. Therefore, the results in \cite{Kdisc} for constructions of F-theory models on genus-one fibered Calabi--Yau 4-folds over base del Pezzo 3-folds generalize to genus-one fibered Calabi--Yau 4-folds over any base 3-folds for 4D F-theory with a discrete $\Z_4$ gauge group. For a discrete $\Z_5$ gauge group, the results in \cite{Kdisc} extend to models over Fano 3-folds as a base, which are more general than del Pezzo 3-folds. 
\par Constructing 6D and 4D F-theory models with a discrete $\Z_5$ gauge group over completely general base spaces remains an open problem. The question of whether the constructions of F-theory models with a discrete $\Z_5$ gauge group discussed in this study generalize to a wider class of base geometries, is a likely target for future studies.
\par The discrete gauge groups that we discussed in this note arise in F-theory models when generic forms of defining equations are chosen for the compactification geometries. When the defining equation takes a special form, it can occur that an $n$-section splits into a multisection of a smaller degree. In this special situation, a discrete gauge group of a smaller degree arises. Examples of this phenomenon is discussed in Section \ref{ssec4.2}, in which F-theory models with a discrete $\Z_2$ gauge group transition to F-theory models with a discrete $\Z_4$ gauge group. This phenomenon physically appears unnatural: as we discussed in section \ref{ssec4.2}, it seems natural to expect that an F-theory model transitions to another F-theory model in which a discrete gauge group is breaking down to another discrete gauge group of a smaller degree, however, a discrete $\Z_2$ gauge group appears to be rather ``enhanced'' to a discrete $\Z_4$ gauge group in the examples of the phenomenon observed in section \ref{ssec4.2}. Studying physical interpretation of this further is a likely direction of future studies.

\section*{Acknowledgments}

We would like to thank Shun'ya Mizoguchi and Shigeru Mukai for discussions.

\end{document}